\documentclass[prd,preprint,%
tightenlines,showpacs,amsfonts]{revtex4}

\begin{document}

\title{Summary of session A4 at the GRG18 conference:\\
Alternative Theories of Gravity}

\author{Gilles \surname{Esposito-Far\`ese}}

\affiliation{${\mathcal{G}}{\mathbb{R}}
\varepsilon{\mathbb{C}}{\mathcal{O}}$, Institut d'Astrophysique
de Paris, UMR 7095-CNRS, Universit\'e Pierre et Marie
Curie-Paris6, 98bis boulevard Arago, F-75014 Paris, France}

\begin{abstract}
More than 50 abstracts were submitted to the A4 session on
\textit{Alternatives Theories of Gravity} at the GRG18 conference.
About 30 of them were scheduled as oral presentations, that
we summarize below. We do not intend to give a critical review,
but rather pointers to the corresponding papers. The main topics
were (i)~brane models both from the mathematical and the
phenomenological viewpoints; (ii)~Einstein-Gauss-Bonnet
gravity in higher dimensions or coupled to a scalar field;
(iii)~modified Newtonian dynamics (MOND); (iv)~scalar-tensor
and $f(R)$ theories; (v)~alternative models involving Lorentz
violations, noncommutative spacetimes or Chern-Simons corrections.
\end{abstract}

\date{November 1, 2007}

\maketitle
\section{Extradimensional theories}\label{5D}

\subsection{Brane models}\label{Branes}
Many recent papers have discussed about the absence or existence of a
ghost degree of freedom in the DGP brane model \cite{Dvali:2000hr}.
K.~Izumi presented his work \cite{Izumi:2006ca} in collaboration with
K.~Koyama and T.~Tanaka, in which the consequences of introducing a
second brane are analysed. It allows to avoid the ghost present in the
helicity-0 component of the massive spin-2 graviton, but a spin-0
ghost is generated instead, and the authors show how they are related.
However, they argue that such a ghost mode may be harmless in de Sitter
background, using the fact that the UV cutoff of the effective theory
is lower than the Planck scale.

Several presentations were devoted to the Randall-Sundrum brane model
\cite{Randall:1999ee,Randall:1999vf}, and more precisely to the second
one (RS-II). It has been conjectured that no brane-localized static
black hole exists in this model if its size is larger than the bulk
curvature scale \cite{Tanaka:2002rb,Emparan:2002px}. N.~Tanahashi,
again in collaboration with T.~Tanaka~\cite{Tanahashi}, checked this
conjecture by constructing time-symmetric initial data, assuming
axisymmetry around the extra dimension. Such initial data can be
constructed for a size of the apparent horizon much larger than the
bulk curvature scale, but the horizon area is always smaller than that
of the black string of the same mass, consistently with the conjecture.
The authors also successfully checked their construction by applying it
to the case of a 4-dimensional bulk spacetime, in which a static
brane-localized black hole solution is known analytically
\cite{Emparan:1999wa}.

A.~Majumdar clearly summarized his works on black holes in
extradimensional theories, and notably in the RS-II model
\cite{Majumdar:2002mra,Majumdar:2003sf,Majumdar:2005tq,Majumdar:2005ba,Mukherjee:2006ru}.
Primordial black holes survive longer in this model (contributing
thus to dark matter), and the author shows that more black-hole
binaries are formed (providing sources of gravitational waves).
Gravitational lensing also differs from the Schwarzschild case,
although it is not yet possible to test these theories with present
experimental accuracy.

L.~Gergely gave a very nice talk about his work \cite{Gergely:2007iw}
in collaboration with Z.~Keresztes and G.~Szab\'o, in which the
cosmological predictions of the RS-II model are confronted with
available supernova data. The authors derive difficult analytical
results, and show that experimental data can be perfectly fitted
within this scenario. However, they insist on the fact that
alternative models are not ruled out either.

Four talks \cite{Kolos,Schee,Kotrlova,Hladik} were given by
students and collaborators of Z.~Stuchl\'{\i}k, who probably holds
the record number of independent abstracts submitted to the
GRG18 conference. There were devoted to various observable effects
of brany black holes, which involve a tidal charge (similar to the
square electric charge $Q^2$ of the standard Reissner-Nordstr\"om
solution, but here with any sign). M.~Kolo\v{s} \cite{Kolos} focused
on circular geodesics around such black holes, and underlined the
quantitative differences with the Reissner-Nordstr\"om case.
J.~Schee \cite{Schee} studied null geodesics around a rotating
brany black hole, i.e., Kerr-Newman geometry with any sign for the
tidal charge. He clearly classified such geodesics in 7 different
regions, and exhibited the small predicted effects on the black hole
images. A.~Kotrlov\'a \cite{Kotrlova} analysed the effect of the tidal
charge on the quasiperiodic oscillations (QPOs) observed in rotating
black holes, and discussed in which situations the spin and the tidal
charge can be determined from observation. She pointed out however that
some systems involving more than one resonant point cannot be explained
even within this brane framework. J.~Hlad\'{\i}k \cite{Hladik} focused
on trapped null geodesics around compact stars, whose exterior
spacetime corresponds again to the above brany black holes. He
studied neutrino trapping in this braneworld as compared to the
general relativistic case \cite{Stuchlik:2007dc}, and showed that
it can be efficient under some precise conditions (but the
experimental accuracy needed to discriminate the theories was not
computed).

As opposed to the above phenomenological studies, the work presented
by J.~E.~Rojas-Marcial \cite{Capovilla:2006dc} was quite mathematical.
He considered Born-Infeld-type theories, in which the determinant
defining the action involves either gauge fields (as usual) or
extrinsic curvature terms related to a brane. The authors constructed
the conserved stress tensor and derived the general field equations.
Because of the presence of higher derivatives, the stability of such
models is however unclear.

\subsection{Gauss-Bonnet}\label{GB}
Adding the Gauss-Bonnet topological invariant $R_{\mu\nu\rho\sigma}^2
- 4 R_{\mu\nu}^2 + R^2$ to the Einstein-Hilbert Lagrangian is well
known not to contribute to the field equations in 4 dimensions, but it
does in higher dimensions or if it is multiplied by a nonconstant
field.

N.~Dadhich and K.~Maeda found in \cite{Maeda:2006iw,Maeda:2006hj} a
static black hole solution in more than 4 dimensions, when both the
Gauss-Bonnet term and a cosmological constant are present (and their
coefficients related). N.~Dadhich presented the recent extension
and interpretation of this solution proposed by the authors
\cite{Dadhich:2007xf}. Pure gravity in 6 dimensions can be seen in 4
dimensions as the collapse of charged null dust forming a black hole.
The authors' conclusion is that matter (the back hole) is produced
by gravity in this case. The audience pointed out that this is not
the first example, since dimensional reduction does generate gauge and
scalar fields from a pure extradimensional metric, and since ingoing
gravitational waves can form a black hole even in 4 dimensions. But
the exact solution \cite{Maeda:2006iw,Maeda:2006hj} remains nice and
interesting anyway.

J.~Oliva presented his work \cite{Dotti:2006cp,Dotti:2007az} in
collaboration with G.~Dotti and R.~Troncoso, in which a static
wormhole solution is constructed in 5-dimensional
Einstein-Gauss-Bonnet gravity with a cosmological constant.
The authors provide a physical interpretation, analyse the causal
structure, and above all prove that the solution is stable against
scalar field perturbations (no symmetry being assumed for such
perturbations).

I.~Neupane, in collaboration B.~Leith \cite{Leith:2007bu},
considered the Gauss-Bonnet term in 4 dimensions but multiplied by a
function of a scalar field. He showed that the dark energy responsible
for the accelerated expansion of the Universe can admit an equation of
state $p/\rho \equiv w < -1$ in this framework, but that it tends to
$-1$ when $t\rightarrow \infty$. Contrary to other models with $w <
-1$, the present one is stable at least at quadratic order (since the
scalar-Gauss-Bonnet term is perturbatively of cubic order), and it
admits nonsingular solutions for a wide range of the
scalar-Gauss-Bonnet coupling constant.

\section{Modified Newtonian dynamics}\label{MOND}

In order to explain the flat rotation curves of galaxies and clusters,
one may invoke either the existence of dark matter or a MOdification of
Newtonian Dynamics at large distances (MOND), as M.~Milgrom proposed
in 1983 \cite{Milgrom:1983ca}. Instead of its Newtonian expression
$a_N = GM/r^2$, the acceleration of a test particle is assumed to read
$a = \sqrt{a_N a_0}$ when $a$ is smaller than a universal constant
$a_0$. This simple recipe superbly accounts for galaxy rotation curves.

However, cluster rotation curves still need a significant amount of
dark matter even within this MOND framework. It has been argued that
neutrinos could form this dark matter, since they can cluster at such
scales (whereas they are too light to cluster at the scale of a
galaxy). R.~Takahashi presented his work in collaboration with T.~Chiba
\cite{Takahashi:2007nj}. They studied weak lensing in three clusters,
and proved that the neutrino mass should be larger that 2 eV to
explain the data (and even $\sim 8$ eV for one cluster). This is thus
inconsistent with the particle physics limit, and the authors'
conclusion is that dark matter around galaxy clusters must involve
something else than neutrinos.

The success of MOND for galaxy rotation curves remains anyway
impressive, and many physicists have thus looked for field theories
reproducing its behaviour as naturally as possible. One of the best
proposals is the Tensor-Vector-Scalar (TeVeS) model constructed by
J.~Bekenstein \cite{Bekenstein:2004ne}. The physical metric to which
matter is assumed to be universally coupled takes however the rather
complicated form $\tilde g_{\mu\nu} \equiv e^{-2\varphi} g^*_{\mu\nu}
- 2 U_\mu U_\nu \sinh(2\varphi)$, where $g^*_{\mu\nu}$ is the Einstein
(spin-2) metric, $U_\mu$ is a unit norm vector field, and $\varphi$ is
a scalar field. M.~Sakellariadou presented her work
\cite{Mavromatos:2007sp} in collaboration with M.~Mavromatos, in which
such fields and such a physical metric are shown to appear naturally in
string theory. The unit vector field comes from the average recoil
4-velocity of D0-branes (point particles) interacting with neutrino
string matter. Moreover, neutrinos not only contribute to the dark
matter density within this scenario, but also to the dark energy
density. Such an increased value of $\Omega_\Lambda$ is precisely
needed in TeVeS to reproduce the CMB spectrum.

G.~Esposito-Far\`ese presented his work in collaboration with
J.-P.~Bruneton \cite{Bruneton:2007si}, in which the stability and the
mathematical consistency of various MOND-like field theories is
analysed, including TeVeS and new models proposed by the authors.
Their conclusion is that no present theory passes all experimental
tests while being stable and admitting a well-posed Cauchy problem,
besides the unnatural fine tuning needed to construct them. On the
other hand, it is possible to account for the Pioneer anomaly (if
confirmed) within a stable and consistent model, but it is fine-tuned
too.

Although the aim of MOND was \textit{a priori} to avoid the dark
matter hypothesis, L.~Blanchet proposed an original reinterpretation
\cite{Blanchet:2006sc,Blanchet:2006yt}: Its phenomenology may be
caused by some special kind of dark matter filling the Universe.
Assuming that it is formed by gravitational dipoles, he shows that
their polarization aligns with the gravitational field of ordinary
matter. Their own gravitational field thereby modifies the Newtonian
one. By choosing an appropriate force law between the gravitational
charges constituting each dipole, the author shows that the MOND
law $a = \sqrt{a_N a_0}$ can be recovered naturally for small
accelerations. The classical version of this model needs to assume the
existence of negative gravitational charges, but the author has also
devised a relativistic version avoiding them.

M.~Seifert presented his study \cite{Seifert:2007fr}, based on his
previous work \cite{Seifert:2006kv} in collaboration with R.~Wald.
He devised a general method to analyse the stability of static,
spherically symmetric solutions in any diffeomorphism covariant field
theory: He considers spherically symmetric perturbations and computes
the eigenvalues of the oscillation frequencies. He then applied this
method to various alternative theories of gravity, and notably
confirmed that the above TeVeS model is unstable. Indeed, the Sun
would not last longer than 2 weeks within this theory. Since he also
applied his method to $f(R)$ and scalar-tensor theories, this provides
a natural transition with the next Section.

\section{Scalar-tensor theories}\label{Scalar-tensor}

\subsection{Brans-Dicke-like models}\label{BD}
Numerical simulations of structure formation are remarkably
successful when assuming the existence of pressureless cold dark
matter, but they generically predict cuspy density profiles at
the center of galaxies. J.~Cervantes-Cota presented his recent
study \cite{CervantesCota} in collaboration with
M.~Rodr\'{\i}guez-Meza and D.~Nu\~nez, in which a
scalar-tensor theory is invoked to avoid such cuspy
profiles. Besides the presence of dark matter, this model
involves two free parameters (the mass of the scalar field and
its coupling constant to matter), that the authors constrain by
fitting both rotation curves and luminosity profiles. They show
that some parameters correspond to shallow density
profiles in galaxy centers. However, the authors do not
discuss whether post-Newtonian tests are passed in
the solar system and binary pulsars. This may be the case
if the scalar field is only coupled to dark matter but not
to baryonic matter. On the other hand, the baryonic
matter-scalar coupling constant they need in
Ref.~\cite{CervantesCota:2007ys}, for instance, is much
too large to satisfy such post-Newtonian constraints.

Scalar-tensor theories can be written in terms of different
variables. The most popular ones are (i)~the Einstein frame, in
which the kinetic term of the metric is the standard
Einstein-Hilbert (spin-2) action, but where the matter action
involves an explicit dependence on the scalar field; (ii)~the
Jordan frame, in which the metric is minimally coupled to matter,
but where the spin-2 and spin-0 kinetic terms are not
diagonalized. The simplest way to prove that the Cauchy problem
is well posed for such theories is to analyse it in the
Einstein-frame, since they reduce then to general relativity, the
scalar field playing the role of an extra matter field. The work
\cite{Salgado:2005hx} that M.~Salgado presented analyses this
Cauchy problem in the Jordan frame instead. This is much more
complicated than in the Einstein frame, and the author needs to
introduce a new class of harmonic slicings to evolve the lapse
function in the $3+1$ ADM decomposition. This allows him to
confirm that the Cauchy problem is indeed well posed. Since the
change of variables from one frame to the other can be singular
in particular situations, the audience was surprized that this
analysis in the Jordan frame did not impose constraints on the
derivatives of the scalar field.

Theories whose Lagrangian involves a function of the scalar
curvature, $f(R)$, are well-known to be equivalent to
scalar-tensor theories. The $R^n$ models studied by J.~Leach, in
collaboration with N.~Goheer and P.~Dunsby \cite{Goheer:2007wu},
are thus of this kind. In the same spirit as their previous paper
\cite{Carloni:2007eu}, they have performed a detailed analysis of
the cosmological dynamics of such models, for anisotropic
homogeneous Bianchi universes. They notably find the nature of
equilibrium points, and investigate static solutions and possible
bouncing behaviours.

\subsection{$f(R)$ theories in the Palatini formalism}\label{f(R)}
The above equivalence of $f(R)$ theories with scalar-tensor
theories assumed that the scalar curvature was a function of the
metric tensor and its first derivatives. There is a subtle but
crucial difference when considering such theories in the first
order (Palatini) formalism, in which the metric $g_{\mu\nu}$ and
the connection $\Gamma^\lambda_{\mu\nu}$ are treated as
independent fields. As shown notably in
\cite{Flanagan:2003rb,Flanagan:2003iw}, $f(R)$ theories are still
equivalent to scalar-tensor theories, but the scalar field does
not propagate: It does not have any kinetic term in the Einstein
frame. It behaves thus as a Lagrange parameter, and its field
equation imposes a constraint. More precisely, its value can be
written in terms of the matter fields, and its presence in the
matter field equations yields thus new unexpected matter-matter
interactions. \'E.~Flanagan showed in 2003
\cite{Flanagan:2003rb,Flanagan:2003iw} that specific models are
ruled out by particle physics data, and that this difficulty is
\textit{a priori} generic for $f(R)$ theories in the Palatini
formalism. [Note however that in these references, the matter
action was assumed to depend on the metric and its derivatives,
but not on the independent connection $\Gamma^\lambda_{\mu\nu}$;
if the first-order formalism is extended to the matter action
itself, the analysis is more involved and the conclusion may be
less dangerous for this class of models.]

E.~Barausse presented his work \cite{Barausse:2007pn}, in
collaboration with T.~Sotiriou and J.~Miller, in which another
serious difficulty of Palatini $f(R)$ theories is pointed out.
They consider polytropic spheres in this framework, and show that
the field equations are singular at their surfaces (because terms
coming from radial derivatives of the function $f'$ behave as
derivatives of the matter density). For a polytropic index
$\frac{3}{2} < \Gamma < 2$, they prove that there is a curvature
singularity at the surface of the body. Their work was criticized
in \cite{Kainulainen:2007bt}, where it is argued that a single
polytrope does not describe realistic stars up to their surface.
However, E.~Barausse underlined that even if one considered a
smooth crust at the surface, this class of theories would anyway
predict huge tidal effects. Large tidal forces would notably
occur in the outer regions of a dilute gas of solar-system size,
and this seems unrealistic.

Independently of the above difficulties, R.~Tavakol analysed in
depth the cosmological dynamics of such Palatini $f(R)$ theories,
in collaboration with S.~Fay and S.~Tsujikawa \cite{Fay:2007gg}.
His presentation was particularly pedagogical, and his main
conclusion was that such theories can predict 3 out of the 4
required cosmological phases (early inflation, radiation
domination, matter domination, late de Sitter expansion). This is
thus better than $f(R)$ theories in the (second order) metric
formalism. However, it seems impossible to predict inflation
\textit{followed} by a radiation-dominated era in the present Palatini
framework.

Still in this framework, K.~Uddin presented his study of
cosmological perturbations \cite{Uddin:2007gj}, in collaboration
with J.~Lidsey and R.~Tavakol. He compared two different
approaches developed in the literature: direct linearization of
the field equations, and an application of Birkhoff's theorem. It
happens that the latter (simpler) method does not yield pressure
gradients in the perturbation equations, contrary to the latter
(more rigorous) one. The authors determine for which theories and
under which conditions the simpler method anyway gives reasonable
results.

\section{Alternative models}\label{Alternative}

Although Lorentz symmetry is very well tested experimentally, it
may happen that it is violated at the Planck scale, where quantum
gravity effects cannot be neglected. R.~Bluhm presented his work
\cite{Bluhm:2004ep}, in collaboration with A.~Kostelecky, in
which they study gravitational effects occurring when Lorentz
symmetry is spontaneously broken. He notably underlined that this
also causes a breaking of diffeomorphism invariance, and the
generation of both massless (Nambu-Goldstone) and massive modes.
The related observable effects are described in terms of an
effective field theory.

A.~Kobakhidze summarized his works with X.~Calmet
\cite{Calmet:2005qm,Calmet:2006iz}, in which they formulate a
theory of gravitation on a noncommutative spacetime. They exhibit
different implementations of general coordinate invariance and
local Lorentz invariance, focusing on unimodular gravity
(volume-preserving diffeomorphisms). They construct twisted gauge
symmetries in the spirit of J.~Wess \textit{et al.}
\cite{Aschieri:2006ye,Fiore:2007vg}, and show how a fully
covariant theory of gravity may be defined on Riemann-Fedosov
manifolds.

Chern-Simons modified gravity was first constructed by S.~Deser
\textit{et al.} in $2+1$ dimensions \cite{Deser:1981wh}, and it
has been extended to $3+1$ dimensions by R.~Jackiw and S.~Pi
\cite{Jackiw:2003pm}. In this theory, the Schwarzschild solution
holds but not the Kerr one. In $2+1$ dimensions, the latter is
replaced by the rotating black hole solution of
Ref.~\cite{Moussa:2003fc}, in which frame dragging is enhanced.
K.~Konno presented his work \cite{Konno:2007ze} in collaboration
with T.~Matsuyama and S.~Tanda, in which they look for rotating
black holes in $3+1$-dimensional Chern-Simons modified gravity.
They do not find an exact solution, but analyse perturbations of
the Schwarzschild one. The theory depends on a fixed vector field
$\nu_\mu$, called the embedding coordinate. When it is timelike,
the authors show that no finite-mass rotating black hole exists
in this theory (at least at the perturbative level). On the other
hand, when $\nu_\mu$ is spacelike, rotating black holes exist for
any mass, and the authors exhibit the form of the solution.

\section{Concluding remarks}\label{Concl}
The main puzzles of theoretical and experimental gravity trigger
our interest for alternative theories. Quantum gravity suggests
possible Lorentz violations, noncommutative spacetimes, or
extradimensional theories. The latter imply the existence of
scalar partners to the graviton, and even more subtle
phenomenological effects occur if our 4-dimensional spacetime is
a brane embedded in the higher-dimensional bulk. On the other
hand, the experimental evidence for 72~\% of dark energy and
24~\% of dark matter in our Universe may be a hint that general
relativity fails at large scales. All the works presented in this
session addressed at least one of these questions. Some no-go
theorems or negative results usefully reduce the space of allowed
models to be explored, while new ideas recall us that Nature
might be richer than we expect. Let us bet that the quest for
alternative theories of gravity will not end until an elegant and
consistent model predicts all experimental data.



\end{document}